\documentclass[prd,nofootinbib,preprint,superscriptaddress]{revtex4-1}

\usepackage{amsmath, amssymb, amsthm, graphicx, epsfig, fancyhdr,epsfig}
\usepackage{amssymb}
\usepackage{tikz-feynman}
\tikzfeynmanset{compat=1.1.0}

\usepackage{graphicx}
\usepackage{dcolumn}
\usepackage{bm}
\usepackage{geometry}
\usepackage{wrapfig}
\usepackage{mathtools}
\usepackage{amsmath}
\usepackage{amsfonts}
\usepackage{mathrsfs}
\usepackage{subcaption}
\usepackage{hyperref}
\usepackage{multirow}
\usepackage{color}
\usepackage{tikzsymbols}
\usepackage{natbib}

\newcommand\ChangeRT[1]{\noalign{\hrule height #1}}

\def\bea{\begin{eqnarray}}
\def\eea{\end{eqnarray}}

\def\beq{\begin{equation}}
\def\eeq{\end{equation}}

\begin{document}
\title{Neutrino mass and $(g-2)_{\mu}$ with dark $U(1)_D$ symmetry}
\author{Arnab Dasgupta}
\email{arnabdasgupta@protonmail.ch}
\affiliation{Institute of Convergence Fundamental Studies , Seoul-Tech, Seoul 139-743, Korea}

\author{Sin Kyu Kang}
\email{skkang@seoultech.ac.kr}
\affiliation{School of Liberal Arts , Seoul-Tech, Seoul 139-743, Korea}

\author{Myeonghun Park}
\email{parc.seoultech@seoultech.ac.kr}
\affiliation{School of Liberal Arts , Seoul-Tech, Seoul 139-743, Korea}

\begin{abstract}
We propose an extension of the Standard model (SM)  for radiative neutrino mass by introducing a dark $U(1)_D$ gauge symmetry. The kinetic mixing between the SM gauges and the dark $U(1)_D$ gauge arises at 1-loop mediated by new inert scalar fields.
We show that the tiny neutrino mass and dark matter candidates are naturally accommodated.
Motivated by the recent measurement of $(g-2)_{\mu}$ indicating $4.2~ \sigma$ deviation from the SM prediction,
we examine how the deviation $\Delta a_{\mu}$ can be explained in this model.
\end{abstract}
\maketitle

\section{Introduction}
The recent measurement of the muon anomalous magnetic moment, $a_\mu$ = $(g - 2)_\mu/2$, by the E989
experiment at Fermilab \cite{Abi:2021gix} shows a discrepancy with respect to the theoretical prediction of the Standard
Model (SM) \cite{Aoyama:2020ynm}
\begin{eqnarray}
a^{\rm FNAL}_\mu = 116 592 040(54) \times 10^{-11}\\
a^{\rm SM}_\mu = 116 591 810(43) \times 10^{-11}
\end{eqnarray}
which when combined with the previous Brookhaven determination \cite{Bennett:2004pv} of
\begin{equation}
a^{\rm BNL}_\mu = 116 592 089(63) \times 10^{-11}
\end{equation}
leads to a 4.2 $\sigma$ observed excess of
$\Delta a_\mu = 251(59) \times 10^{-11}$.
The status of the SM calculation of the muon anomalous magnetic moment $(g-2)_{\mu}$  has been updated recently in \cite{Aoyama:2020ynm, Zyla:2020zbs}, although  the hadronic contributions are still challenging to reliably estimate \cite{Blum:2013xva}.
The apparent discrepancy with a $4.2~\sigma$ deviation between 
the SM prediction and the measured value of $(g-2)_{\mu}$ is a long-standing puzzle which may point to new physics.
Inspired by  the latest Fermilab measurements, several literatures that explains the discrepancy with updating possible theoretical models
\cite{Arcadi:2021cwg,Zhu:2021vlz,Han:2021gfu, Baum:2021qzx, Bai:2021bau, Das:2021zea,
Ge:2021cjz, Brdar:2021pla, Buen-Abad:2021fwq,
Zu:2021odn, Amaral:2021rzw,
Ahmed:2021htr, Abdughani:2021pdc, VanBeekveld:2021tgn, Cox:2021gqq, Wang:2021bcx, Gu:2021mjd, Cao:2021tuh, Yin:2021mls, Han:2021ify, Aboubrahim:2021rwz, Yang:2021duj, Ferreira:2021gke, Wang:2021fkn, Li:2021poy, Cadeddu:2021dqx, Calibbi:2021qto, Chen:2021vzk, Escribano:2021css, Chun:2021dwx,
Athron:2021iuf}.

In this paper,  we study how the recent measurement of $(g-2)_{\mu}$ can be explained in a model for radiative generation of tiny neutrino masses. We extend the scotogenic model by introducing a dark $U(1)_D$ gauge symmetry.
The spontaneous symmetry breaking of dark $U(1)_D$ for a massive a ``dark photon" induces $\mathbb{Z }_2$ symmetry, an essential ingredient in scotogenic scenarios for tiny neutrino masses through 1-loop contribution.
The neutral component of a new inert scalar doublet is responsible for dark matter (DM).
In this model, kinetic mixing between the SM neutral gauge bosons and
the dark photon is naturally generated through 1-loop mediated by new scalar fields.
The introduction of a few new particles can reconcile the discrepancy between the recent measurement and the SM prediction. 
In this paper, we will examine the contributions of new scalars and dark photon to $(g-2)_{\mu}$ and see how
they can accommodate the $4.2~\sigma$ deviation of $(g-2)_{\mu}$.

This paper is organized as follows. In section II, we present the setup and the content of new particles by assigning quantum numbers appropriately. In section III, we show how neutrino mass can arise from 1-loop
and present the kinetic mixing between SM gauge bosons and the dark photon.
The contributions of new scalars and the dark photon to $(g-2)_{\mu}$ are explicitly shown. In section IV, numerical results and discussion are presented. We finally conclude in section V.

\section{Model}
To generate tiny neutrino masses radiatively and to have natural DM candidates, we take the framework
of the scotogenic model where two scalar doublets $(\eta_1, \,\eta_2)$, two scalar singlets $(\phi,\, \chi)$ and two vector-like neutral fermions $\Psi_{L,R}$ are introduced.
The complete content of the new fields introduced and their quantum number assignments under $SU(2)_L\otimes U(1)_Y\otimes U(1)_D$ gauge symmetry are shown in table~\ref{tab:U1cont1}.
We note that the model is anomaly free.
\begin{table}
\caption{New particle content with quantum number assignment under an extended gauge symmetry. }
\label{tab:U1cont1}
\centering
\begin{tabular}{!{\vrule width 1.1pt}c|c|c|c|c!{\vrule width 1.1pt}}
\ChangeRT{1.1pt}
Field & SU(2)$_L$ & U(1)$_Y$ & U(1)$_D$ & 2S+1 \\ \ChangeRT{1.1pt}
$\eta_1$  & \textbf{2} &  $\frac{\textbf{1}}{\textbf{2}}$ &  \textbf{-1}& 1 \\ \hline
$\eta_2$  & \textbf{2} &  $\frac{\textbf{1}}{\textbf{2}}$ &  \textbf{1}& 1 \\ \hline
$\phi$  & \textbf{1} &  \textbf{0} &  \textbf{1}&  1 \\ \hline
$\chi$ &  \textbf{1} &  \textbf{0} &  \textbf{2}& 1 \\ \hline
$\Psi_L$ &  \textbf{1} & \textbf{0} &  \textbf{1}& 2 \\ \hline
$\Psi_R$ &  \textbf{1} &\textbf{0} &  \textbf{-1}&2 \\ \ChangeRT{1.1pt}
\end{tabular}
\end{table}
The Lagrangian for the new fermions is given by
\bea\label{IRHYukawa}
{\cal L_F}  \supset &&    y_{1\nu} \overline{L} \tilde{\eta_1}\Psi_R + y_{2\nu} \overline{\Psi^c_L} \tilde{\eta_2}L+y_{\Psi_L}\chi^{\ast}\overline{\Psi}^c_L\Psi_L
            +y_{\Psi_R}\chi \overline{\Psi}^c_R\Psi_R 
            + M_{\Psi} \overline{\Psi}_L\Psi_R+h.c.  \label{new-fermion}
\eea
The Lagrangian for the entire scalar sector is given by
\bea
{\cal L_S}  \supset   && m^2_H H^{\dagger} H+m^2_{\phi} \phi^{\dagger}\phi +m^2_{\chi} \chi^{\dagger}\chi +m^2_{\eta_1} \eta_1^{\dagger}\eta_1 
 +m^2_{\eta_2} \eta_2^{\dagger}\eta_2+ \tilde{\mu_2} \tilde{\eta_1}\eta_2\chi+\tilde{\mu_1} \chi^{\ast}\phi\phi \nonumber \\
 && + \lambda_H (H^{\dagger} H)^2
 +\lambda_{\phi}(\phi^{\dagger}\phi)^2+\lambda_{\chi}(\chi^{\dagger}\chi)^2  +\lambda_{\eta_1}(\eta_1^{\dagger}\eta_1)^2
+\lambda_{\eta_2}(\eta_2^{\dagger}\eta_2)^2 
 \nonumber \\
&&+ \lambda_{\phi\eta_1}(\phi^{\dagger}\phi)(\eta_1^{\dagger}\eta_1) 
 + \lambda_{\phi\eta_2}(\phi^{\dagger}\phi)(\eta_2^{\dagger}\eta_2)+ \lambda_{\phi\chi}(\phi^{\dagger}\phi)(\chi^{\dagger}\chi) +
 \lambda_{\chi\eta_1}(\chi^{\dagger}\chi)(\eta_1^{\dagger}\eta_1)
 \nonumber \\
&& +  \lambda_{\phi\eta_2}(\phi^{\dagger}\phi)(\eta_2^{\dagger}\eta_2)+ \lambda_{\eta_1\eta_2}(\eta_1^{\dagger}\eta_1)(\eta_2^{\dagger}\eta_2)+\lambda_{\eta_1\eta_2\phi}\tilde{\eta_1}\eta_2\phi^2 
+\lambda_{H\eta_1\phi\chi} H^{\dagger}\eta_1\phi^{\ast} \chi
\nonumber \\
&& +\lambda_{H\eta_2\phi\chi} H^\dagger\eta_2\phi \chi^{\ast}
+ (H^{\dagger}H)(\lambda_{H\eta_1}\eta_1^{\dagger}\eta_1+
\lambda_{H\eta_2}\eta_2^{\dagger}\eta_2+
\lambda_{H\chi}\chi^{\dagger}\chi+
\lambda_{H\phi}\phi^{\dagger}\phi),
\eea
where $H$ represents the SM Higgs scalar whose vacuum expectation value (VEV) is denoted by $v$.
In addition the SM Higgs,  $\chi$ gets the VEV, which spontaneously breaks $U(1)_D$, and then the scalar $\chi$ can be written as
\beq
\chi=v_{\chi}+h_{\chi}+i\xi_{\chi},
\eeq
where $v_{\chi}$ is VEV of $\chi$, and $h_{\chi}, \xi_{\chi}$ denote the real and imagiranry components of the field, respectively.
After Ew symmetry as well as $U(1)_D$ are spontaneously broken, we obtain the squared mass matrix for the scalars $H, \chi$ given as
\begin{align}
M_{H\chi}^2=\left(\begin{matrix}
\lambda_{H}v^2 & \lambda_{H\chi} v v_{\chi} \\
\lambda_{H\chi} v v_{\chi}  & \lambda_{\chi} v^2_{\chi}
\end{matrix}\right).
\end{align}

The squared mass matrix is diagnalized by the mixing matrix defined as
\begin{align}
\left(\begin{matrix}
H^0\\  \chi
\end{matrix}\right)= \left(\begin{matrix}
\cos \theta & \sin \theta \\
-\sin \theta & \cos \theta
\end{matrix}\right)\left(\begin{matrix}
h_1 \\ h_2
\end{matrix}\right),
\end{align}
where the mixing angle $\theta$ is given as
\beq
\tan 2\theta = \frac{2\lambda_{H\chi} v_{\chi}}{\lambda_{\chi}v^2_{\chi}-\lambda_H v^2}
\eeq

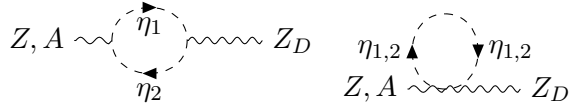
\begin{figure}
    \centering
    \begin{tabular}{lr}
       \begin{tikzpicture}[/tikzfeynman/small]
    \begin{feynman}
    \vertex (i){$Z,A$};
    \vertex[right = 1.cm of i](v1);
    \vertex[right = 1.cm of v1](v2);
    \vertex[right = 1.cm of v2](j){$Z_D$};
    \diagram*[small]{(i)--[boson](v1)--[charged scalar,half left,edge label' = $\eta_1$](v2)--[boson](j),(v2)--[charged scalar,half left,edge label=$\eta_2$](v1)};
    \end{feynman}   
    \end{tikzpicture}  &  
      \begin{tikzpicture}[/tikzfeynman/small]
    \begin{feynman}
    \vertex (i){$Z,A$};
    \vertex[right = 1.cm of i](v1);
    \vertex[above = 1.cm of v1](v2);
    \vertex[right = 1.cm of v1](j){$Z_D$};
    \diagram*[small]{(i)--[boson](v1)--[boson](j),(v1)--[charged scalar,half left,edge label=$\eta_{1,2}$](v2)--[charged scalar,half left,edge label=$\eta_{1,2}$](v1)};
    \end{feynman}   
    \end{tikzpicture}
    \end{tabular}
    \caption{Feynman diagrams for the amplitudes generating the Kinetic mixing.}
    \label{fig:km}
\end{figure}

We take into account the kinetic mixing between the SM gauges and $U(1)_D$ gauge which arises via 1-loop.
Then, the kinetic terms of the gauge fields containing the kinetic mixing are given by
\beq
{\cal L_{\rm kin}}=-\frac{1}{4}\hat{F}_{\mu\nu}\hat{F}^{\mu\nu}-\frac{1}{4}\hat{Z}_{\mu\nu}\hat{Z}^{\mu\nu}-\frac{1}{4}\hat{V}_{\mu\nu}\hat{V}^{\mu\nu}
-\frac{\epsilon}{2}\hat{F}_{\mu\nu}\hat{V}^{\mu\nu}-\frac{\epsilon_{ZV}}{2}\hat{Z}_{\mu\nu}\hat{V}^{\mu\nu},
\eeq
where $\hat{F}_{\mu\nu}, \hat{Z}_{\mu\nu}$ , and $\hat{V}_{\mu\nu}$ represent the gauge field strengths for $U(1)_Y, SU(2)_L$ , and $U(1)_D$, respectively.
The relevent piece of the covariant derivatives is given as
\beq
-i e Q A_{\mu}-i\left[ \frac{g}{c_W}(T^3_L-s^2_W Q)-g_D s_{\theta_{2V}} Q_D\right] Z_{\mu} -i\left[\epsilon e Q+g_D Q_D\right] Z^{\prime}_{\mu}
\eeq
where $Q$ and $Q_D$ are the charges of $U(1)_Y$ and $U(1)_D$, respectively.
The kinetic terms are diagonalized by the field redefinition
given as
\begin{align}
\left(\begin{matrix}
\hat{A}_{\mu} \\ \hat{Z}_{\mu} \\ \hat{V}_{\mu}
\end{matrix}\right)= \left(\begin{matrix}
1 & 0&-\frac{\epsilon}{D} \\
0& 1& -\frac{\epsilon_{ZV}}{D} \\
0 & 0 & \frac{1}{D}
\end{matrix}\right)\left(\begin{matrix}
A_{\mu}\\ Z_{\mu} \\ Z^{\prime}_{\mu}
\end{matrix}\right).
\end{align}
Following Ref.\cite{Rueter:2020qhf}, the mixing parameter $\epsilon$ generated from the 1-loop diagram mediated by two scalars $\eta_1, \eta_2$ as shown in fig.\ref{fig:km} and is given as
\beq
\epsilon = \frac{eg_D}{48\pi^2} {\rm ln} \left(\frac{M^2_{\eta_1}}{M^2_{\eta_2}}\right). \label{epsilon}
\eeq

In this work, we ignore the term proportional to $\epsilon_{ZV}$ which
is in general much smaller than $\epsilon$.

\section{Radiative neutrino mass and New contributions to $(g-2)_{\mu}$}
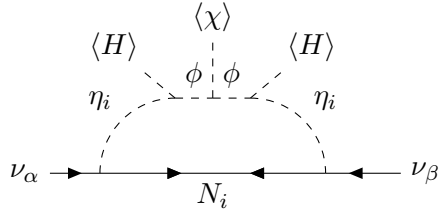
\begin{figure}
    \centering
    \begin{tikzpicture}[/tikzfeynman/small]
    \begin{feynman}
    \vertex (i){$\nu_\alpha$};
    \vertex[right = 1.cm of i](v1);
    \vertex[right = 3.cm of v1](v3);
    \vertex[right = 1.cm of v3](j){$\nu_\beta$};
    \vertex[above right = 1.414 cm of v1](v5);
    \vertex[ right = 0.5 cm of v5](v6);
    \vertex[above left = 1.414 cm of v3](v7);
    \vertex[above left = 0.5 cm of v5](k){$\langle H \rangle$};
    \vertex[above right = 0.5 cm of v7](l){$\langle H \rangle$};
    \vertex[above = 0.75 cm of v6](m){$\langle \chi \rangle$};
    \diagram*[small]{(i)--[fermion](v1)--[majorana,edge label' = $N_i$](v3)--[anti fermion](j),(v1)--[scalar,quarter left,edge label=$\eta_i$](v5)--[scalar,edge label=$\phi$](v6)--[scalar,edge label=$\phi$](v7)--[scalar,quarter left,edge label=$\eta_i$](v3),(v5)--[scalar](k),(v7)--[scalar](l),(v6)--[scalar](m)};
    \end{feynman}   
    \end{tikzpicture}
    \caption{Neutrino Mass production.}
    \label{fig:neu_mass}
\end{figure}

Let us first study how the light neutrino mass can be radiatively generated.
From the field redefinition of the neutral fermions whose mass terms are presented in Eq.(\ref{new-fermion}) given as
\begin{align}
\left(\begin{matrix}
\Psi_L\\  \Psi_R
\end{matrix}\right)= \left(\begin{matrix}
\cos \theta_N & \sin \theta_N \\
-\sin \theta_N & \cos \theta_N
\end{matrix}\right)\left(\begin{matrix}
N_1 \\ N_2
\end{matrix}\right),
\end{align}
we can obtain mass eigenvalues of the new neutral fermions after $\chi$ gets VEV given as
\bea
M_{N_1(2)}=\frac{1}{2}[y_{\Psi_L}v_{\chi}+y_{\Psi_R}v_{\chi}\mp\sqrt{M_{\Psi}^2-4y_{\Psi_L}y_{\Psi_R}v_{\chi}^2 }].
\eea
As for the scalar doublet the mass matrix can be simplified by assuming the $\phi$ only mixes with the lightest doublet hence the mixing matrix takes the following form
\begin{equation}
    m^2_{R,I} = \begin{pmatrix} 
    \mu^2_{\eta_1} & \lambda_{H\eta_1 \phi \chi} v v_{\chi} \\
    \lambda_{H\eta_1\phi \chi} v v_{\chi} & \mu^2_{\phi} \pm \mu_1 v_\chi
    \end{pmatrix}.
\end{equation}
 Consequently, the projection to the mass basis is given as
 \begin{align}
     \eta_{R,I} &= \cos(\theta_{R,I})\xi_{1R,I} + \sin(\theta_{R,I})\xi_{2R,I} \nonumber \\
     \phi_{R,I} &= -\sin(\theta_{R,I})\xi_{1R,I} + \cos(\theta_{R,I}) \xi_{2R,I}
 \end{align}
The 1-loop Feynman diagram for the generation of neutrino mass is shown in Fig. \ref{fig:neu_mass}.
The neutrino mass  is explicitly given by
\begin{align}
\label{eq:mloop}
(M_\nu)_{\alpha \beta}&=\sum_ky_{\alpha k}y_{\beta k}\Lambda_k  \\
\Lambda_k &= \frac{M_{N_k}}{16\pi^2} \left[\cos^2(\theta_R)F(x_{1R}) + \sin^2(\theta_R)F(x_{2R}) \right.
\label{eq:lambda} \nonumber \\
&- \left.\cos^2(\theta_I)F(x_{1I}) - \sin^2(\theta_I)F(x_{2I})\right],
\end{align}
where
$F(x_i,x_j)$ is defined as
\begin{align}
\label{eq:f1}
F(x)&=\frac{x}{x-1}\ln[x],
\end{align}
with $x_{i}=\frac{m_{i}^2}{M_{N_2}^2}$.
\textcolor{red}{} For the purpose of numerical calculation,
We take Casas-Ibarra (CI) parametrisation \cite{Casas:2001sr} 
for the matrix of the Yukawa coupling satisfying the neutrino data given as
\begin{align}
y_{i\alpha} \ = \ \left(U D_\nu^{1/2} R^{\dagger} \Lambda^{1/2} \right)_{i\alpha} \, ,
\label{eq:Yuk}
\end{align}
where $R, U D_{\nu}$ and $\Lambda$ are an arbitrary complex orthogonal matrix,  the neutrino mixing matrix,
diagonal light neutrino mass matrix , and diagonal loop factor given in eq.(\ref{eq:lambda}).
For the sake of simplicity, we take $R$ to be identity. As for $U$ and $D_{\nu}$, we take experimental results  for the case of normal hierarchy given in \cite{Zyla:2020zbs}.

Now, let us consider the contributions of dark sector to $(g-2)_{\mu}$.
In our model, new particles are assumed to be heavy except for the dark photon $Z^{\prime}$ and a light dark scalar $h_2$ whose masses are order of a few 100 MeV.
Then, sizable corrections to $(g-2)_{\mu}$ arises from the interactions \bea
 &(e\epsilon)\overline{\mu}\gamma^{\mu} \mu Z^{\prime}_{\mu}, \nonumber \\
&(\sin \theta y_{\mu})\overline{\mu}\mu h_2
\eea
The $(g-2)_{\mu}$ has new three main contributions which can be categorized as follows:
\begin{enumerate}
    \item Mediated by dark scalar $\chi$ through SM Higgs Dark scalar mixing.
    \item Mediated by the dark photon through kinetic mixing. 
    \item Mediated by the inert-double and the neutral fermion.
\end{enumerate}
The 1-loop contribution mediated by the dark scalar $h_2$ to  $(g-2)_{\mu}$ is given by\cite{Queiroz:2014zfa}.
\bea
\Delta a_{\mu}(h_2)&=&\frac{y^2_{\mu}s^2_{\theta}}{4\pi^2} \frac{m^2_{\mu}}{M^2_{h_2}} \int^{1}_{0} dx
\frac{x^2 (2-x)}{(1-x)(1-\xi_{h_2}^2 x)+\xi_{h_2}^2 x} \nonumber \\
&=&\frac{y^2_{\mu}s^2_{\theta}}{4\pi^2} \frac{m^2_{\mu}}{M^2_{h_2}}\left[ {\rm ln}\left( \frac{M_{h_2}}{m_{\mu}}\right)  -\frac{7}{12} \right],
\eea
where $\xi_{h_2}=m_{\mu}/M_{h_2}$.
The 1-loop contribution mediated by the dark photon $Z^{\prime}$ to  $(g-2)_{\mu}$ is given by
\beq
\Delta a_{\mu}(Z^{\prime}_{\mu})=\frac{e^2\epsilon^2}{8\pi^2} \frac{m^2_{\mu}}{M^2_{Z^{\prime}}} \int^{1}_{0} dx
\frac{2x^2 (1-x)}{(1-x)(1-\xi_{Z^{\prime}}^2 x)+\xi_{Z^{\prime}}^2 x}
\eeq
where $\xi_{Z^{\prime}}=m_{\mu}/M_{Z^{\prime}}$.

Along with these there is another contribution coming from the neutral fermion and the charged scalar which is given as 
\begin{align*}
    \Delta a_\mu(\eta,N) &= \sum_{ik}\frac{y^{\eta_i*}_{\mu k}y^{\eta_i}_{\mu k}}{16\pi^2} \left[\frac{1}{6(\alpha_{ik} - 1)^4}\left(-2\alpha^3_{ik} - 3\alpha^3_{ik} + 6\alpha_{ik} -1+6\alpha^2_{ik}\ln[\alpha_{ik}] \right)\right] ,
\end{align*}
 where $\alpha_{ik} = m^2_{N_k}/m^2_{\eta_i}$.
It is worthwhile to note that the contribution mediated by the dark scalar is negligibly small because the yukawa coupling for the muon is already of the order $10^{-4}$ and small mixing between the dark scalar and the SM Higgs gives another suppression.

\section{Numerical Result and discussion}
Ignoring the contributions from the light dark scalar, 
the total contribution of $\Delta a_{\mu}$ is given by
the addition of the 1-loop contributions mediated by the dark photon, and by inert scalar bosons accompanied by neutral fermions as follows:
\begin{align}
    \Delta a_\mu &= \Delta a_\mu(Z') + \Delta a_\mu(\eta,N).
\end{align}
We note that $\Delta a_\mu(\eta,N)$ has a overall minus sign which ends up contributing negatively and needs to be compensated by the contribution coming from the kinetic mixing $\Delta a_\mu(Z')$.

In our analysis, we assume that the mass of the dark photon
is around 100 MeV, whereas the mass of the lightest inert neutral scalar is around $\geq \mathcal{O}(500)$ GeV which is the favorable regime for the good dark matter candidate.
For the yukawa coupling $y_{\nu}$, we take to be $\mathcal{O}(1)$ so that we can achieve the tiny neutrino mass of order atmospheric mass scale by tuning the trilinear coupling $\mu_i$
small. But we keep the mass splitting between $\eta_{DMR}$ and $\eta_{DMI}$ to be above
$\mathcal{O}$(keV) to circumvent the constraint coming from direct detection resulting from $Z-$ mediation.  
Under the assumptions above, we see that the contribution from the 1-loop mediated by the dark photon is dominant over the others.

\begin{figure}
    \centering
    \begin{tabular}{lcr}
       \includegraphics[width=0.45\textwidth]{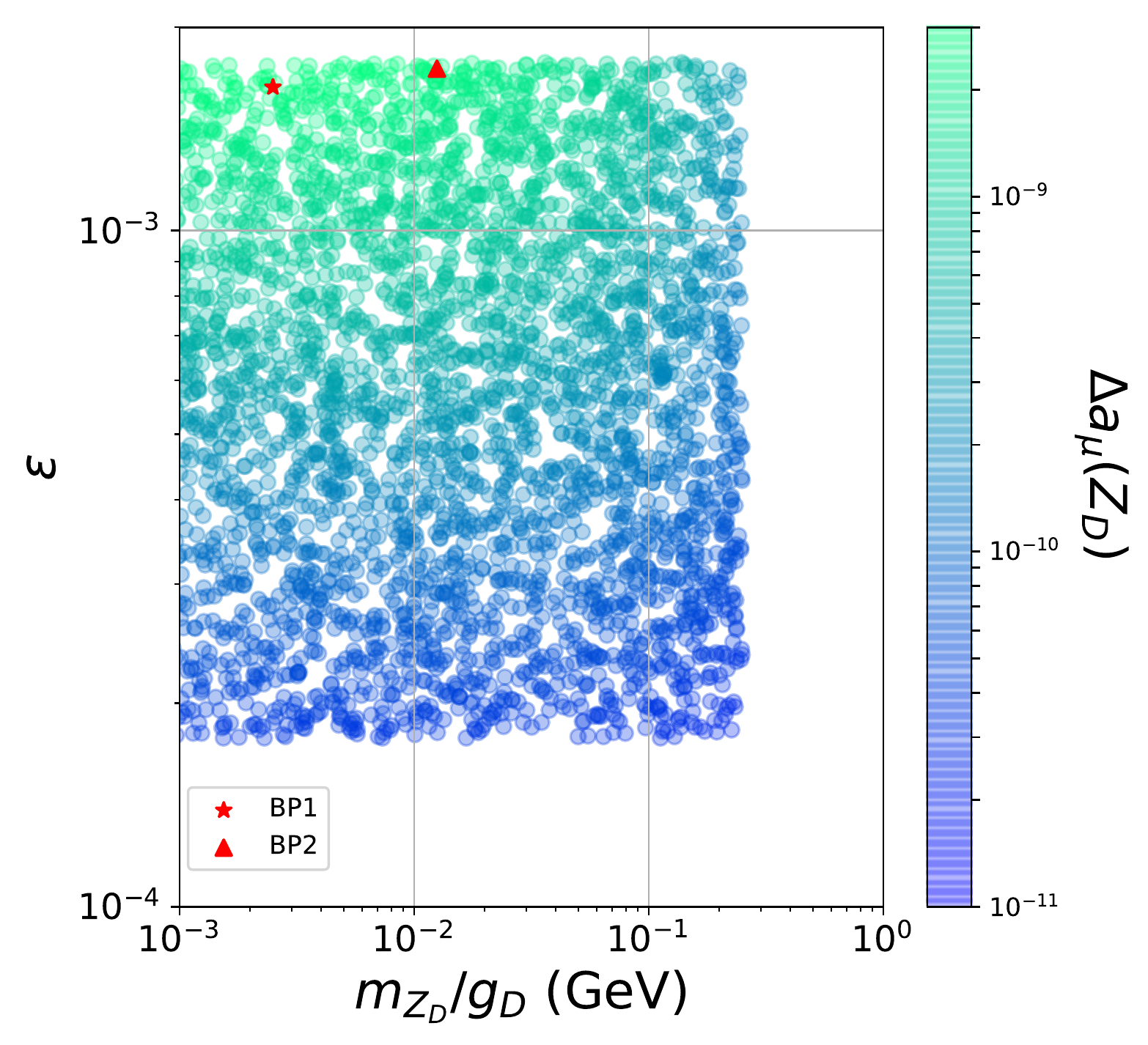} &
       \includegraphics[width=0.45\textwidth]{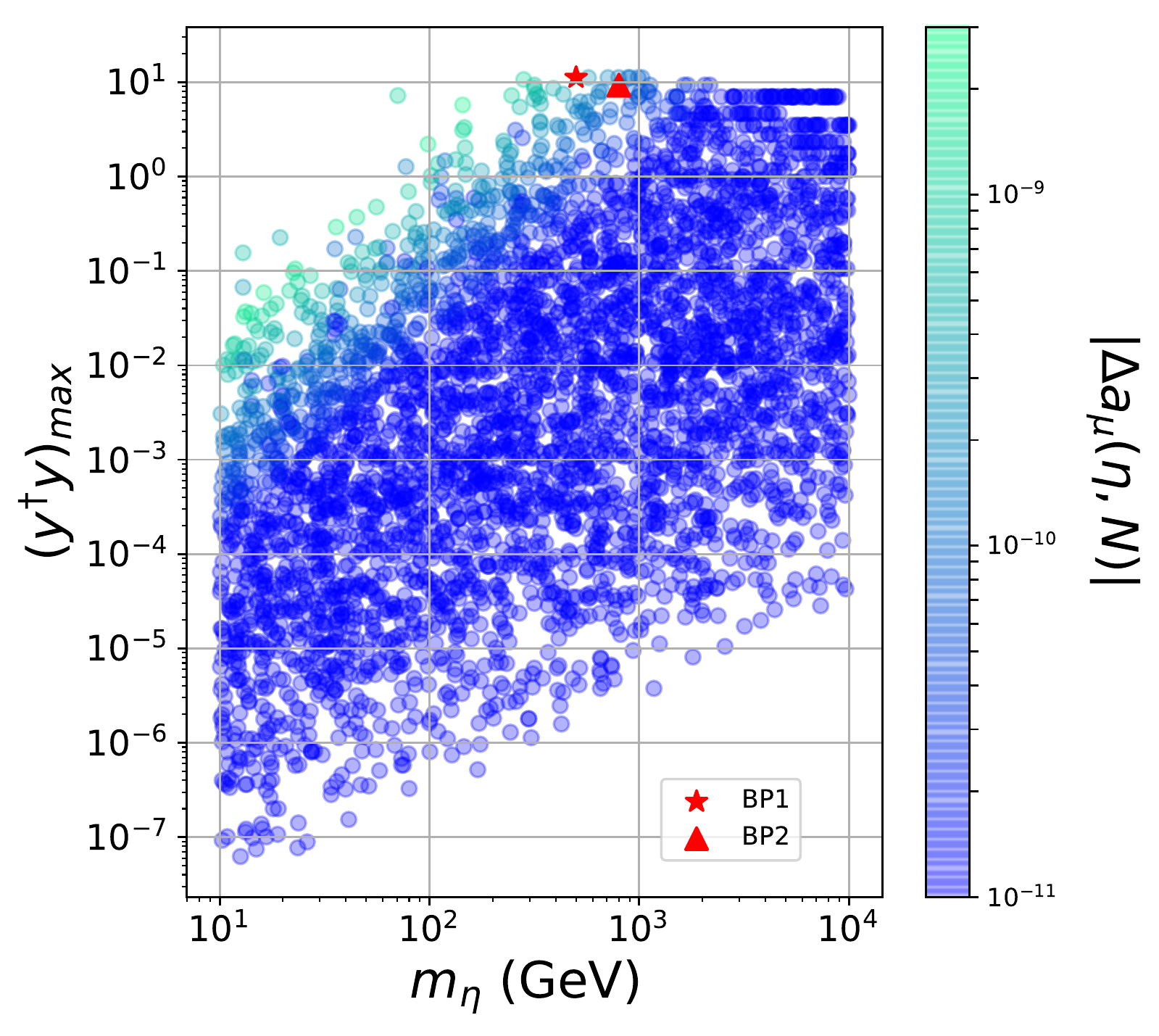} \\
    \end{tabular}
    \includegraphics[width=0.45\textwidth]{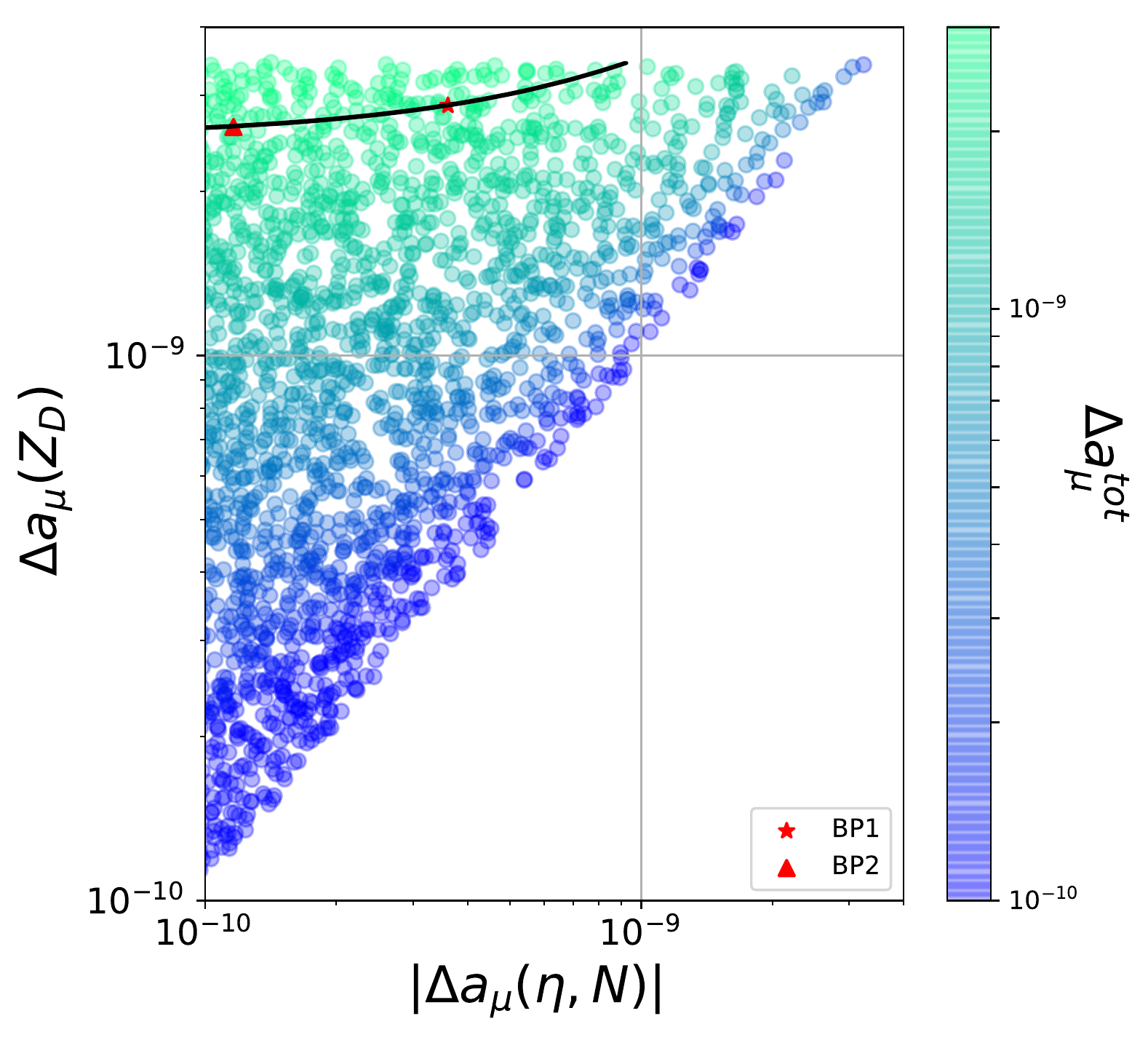}
    \caption{Predictions of $\Delta a_{\mu}(Z^{\prime})$  in terms of $(m_{Z_D}/g_D, \epsilon)$ (upper right) and those of $\Delta a_{\mu}(\eta, N)$  in terms of $(m_{\eta}, y^{\dagger}_{\nu}y_{\nu})$ (upper left).
    Predictions of $\Delta a_{\mu}$ along with $\Delta a_{\mu}(Z^{\prime})$ and $\Delta a_{\mu}(\eta, N)$. Two red points correspond to the two bench mark points presented in Table \ref{tab:my_label} and
    the black curve correspond to $4.2\sigma$ deviation.}
    \label{fig:plots}
\end{figure}

To calculate $\Delta a_{\mu}$ we scan the ranges of the input parameter given as
\begin{align*}
 1 {\rm GeV} &\lesssim m_{\eta_1} \lesssim 1000 {\rm GeV}, \nonumber \\
  0.1 &\lesssim g_D \lesssim 1, \nonumber\\ 
 & m_{\eta_2} = 4 m_{\eta_1}, \nonumber \\
 & m_{\phi} = 5 m_{\eta_1}, \nonumber \\
 & m_{N_i} = m_{\eta_1}. \nonumber 
\end{align*}
Note that the parameter $\epsilon$ can be calculated from eq.(\ref{epsilon}).
In Fig. \ref{fig:plots}, we present how $\Delta a_{\mu}$ is predicted in terms of model parameters.
The upper left panel shows the predictions of $\Delta a_{\mu}(Z^{\prime})$ in the parameter space $(m_{Z_D}/g_D, \epsilon)$, whereas the upper right panel shows  the predictions of $\Delta a_{\mu}(\eta, N)$ in the parameter space $(m_{\eta}, y^{\dagger}_{\nu}y_{\nu})$.
The two red points correspond to the bench mark points presented in Table \ref{tab:my_label}.
The lower panel shows how $\Delta a_{\mu}$ is composed of the two contributions.
As can be seen, $\Delta a_{\mu}(Z^{\prime})$ is dominant over  $\Delta a_{\mu}(\eta, N)$ 
for the ranges of our parameter space we scan.
The black curve in the plot corresponds to the $4.2~ \sigma$ deviation from the SM prediction.
We have checked that the bench mark points present in the Table \ref{tab:my_label} can accommodate the tiny neutrino mass or order of atmospheric mass scale.


\begin{table}[]
    \centering
    \begin{tabular}{|c|c|c|}
    \hline
    Parameters   &   BP1 & BP2  \\ \hline\hline
      $g_D$   & 0.92 &  0.98 \\ \hline
      $v_\chi$ & 10 MeV &  50 MeV \\ \hline
      $\mu_1$  & $1.14$ (keV)& $460.8$ (eV)\\  \hline
      $m_{\eta_1}$ & 500 GeV &  800 GeV \\ \hline
      $m_{\eta_2}$ & 2 TeV &  3.2 TeV \\ \hline
      $m_{\phi}$ & 2.5 TeV &  3.6 TeV\\ \hline
      $\lambda_{H \chi}$ & 0.01 & 0.01 \\ \hline
      $\lambda_{\chi}$ & 0.001& 0.001\\ \hline
      $\lambda_{H \eta_1 \phi}$ & 0.4 & 0.4 \\ \hline 
      $\lambda_{H \eta_2 \phi}$ & 0 & 0\\ 
      \hline
    \end{tabular}
    \caption{Two bench mark points accommodating the tiny neutrino mass around atmospheric scale, dark matter relic density and the $4.2~\sigma$ deviation of $(g-2)_{\mu}$.}
    \label{tab:my_label}
\end{table}

\section{conclusion}
We have considered a dark $U(1)_D$ extension of the SM gauge symmetry to achieve the tiny neutrino mass and to have dark matter candidate.
In this model, the kinetic mixing between the SM gauges and the $U(1)_D$ gauge naturally arises at 1-loop mediated by new inert scalar fields.
The light neutral inert scalar boson can be a good dark matter candidate.
Motivated by the recent measurement of $(g-2)_{\mu}$ indicating $4.2~ \sigma$ deviation from the SM prediction,
we have studied how the deviation $\Delta a_{\mu}$ can be explained in this model.
The 1-loop mediated by dark photon with mass of order 100 MeV as well as inert scalar fields accompanied by the neutral fermions can accommodate the recent measurement of $(g-2)_{\mu}$ at Fermilab.

\bibliographystyle{unsrt}
\bibliography{ref}

\end{document}